# Thin-film radiative thermal diode with large rectification

Qizhang Li[1,2,*], Haiyu He[1,3,*], Qun Chen[2,†] and Bai Song[1,3,4,†]

[1]*Beijing Innovation Center for Engineering Science and Advanced Technology, Peking University, Beijing 100871, China*

[2]*Key Laboratory for Thermal Science and Power Engineering of Ministry of Education, Department of Engineering Mechanics, Tsinghua University, Beijing 100084, China*

[3]*Department of Energy and Resources Engineering, Peking University, Beijing 100871, China*

[4]*Department of Advanced Manufacturing and Robotics, Peking University, Beijing 100871, China*

[†]Corresponding author. Email: songbai@pku.edu.cn, chenqun@tsinghua.edu.cn

[*]These authors contributed equally to this work.

## Abstract

We propose a mechanism to substantially rectify radiative heat flow by matching thin films of metal-to-insulator transition materials and polar dielectrics in the electromagnetic near field. By leveraging the distinct scaling behaviors of the local density of states with film thickness for metals and insulators, we theoretically achieve rectification ratios over 140—a 10-fold improvement over the state of the art—with nanofilms of vanadium dioxide and cubic boron nitride in the parallel-plane geometry at experimentally feasible gap sizes (~100 nm). Our rational design offers relative ease of fabrication, flexible choice of materials, and robustness against deviations from optimal film thicknesses. We expect this work to facilitate the application of thermal diodes in solid-state thermal circuits and energy conversion devices.

Thermal diodes—in analogy to electrical diodes—are two-terminal devices that rectify the flow of heat, and are desired in many applications including thermal management, energy conversion, and thermal circuits and computing [1,2]. The key figure of merit for a thermal diode is its rectification ratio—defined as $R = (Q_F - Q_R)/Q_R$, where $Q_F$ denotes the larger heat current under a forward temperature bias and $Q_R$ the smaller reverse current [3]. A 10-fold rectification is often necessary for technical efficiency. However, despite extensive studies of thermal diodes mediated by phonons, electrons, and photons, most experimentally demonstrated rectification ratios have remained around 1 or much smaller [3–7]. With potential for orders-of-magnitude rectification over a wide range of temperatures, near-field radiative thermal diodes (NFRTD) leveraging the large electromagnetic local density of states (LDOS) close to a surface have recently gained broad interest [6,8-20].

Research on NFRTD began with bulk materials in the parallel-plane geometry separated by a vacuum gap. By using the temperature-dependent surface phonon polaritons (SPhPs) in two polymorphs of silicon carbide (SiC), Otey et al. [8] obtained a rectification ratio of 0.41 at a gap size of 100 nm. Subsequently, the surface modes supported by doped semiconductors [9,17], phase change materials [11,12,14], polar dielectrics [13,16], and two-dimensional materials [15,19] were widely explored together with nanoscale structural innovations (e.g. multilayers and linear gratings), leading sometimes to rectifications over 10. By using nanospheres instead of parallel planes, the largest rectification to date ($>10^4$) has been theoretically predicted [10,18]. However, experiments on NFRTD are rare and have only reported rectifications around 1 [6], largely due to challenges in achieving 10 nm gaps between parallel planes and in charactering radiative thermal transport between delicate nanostructures.

In this letter, we propose that large rectification of radiative heat flow can be readily achieved by matching and optimizing a thin film of metal-to-insulator transition (MIT) materials and another of polar dielectrics in the parallel-plane geometry which is the most technologically promising, at experimentally feasible gap sizes (Fig. 1a). We begin with an asymptotic analysis of hypothetical materials to illustrate the physical mechanism, followed by rigorous calculations for thermal diodes composed of realistic materials and structures. We show over 140-fold rectification near room temperature at around 100 nm gaps, together with a high robustness against deviations from optimal film thicknesses. Our results open up the possibility for realizing orders-of-magnitude thermal rectification using state-of-the-art experimental techniques [21–24].

To design a high-performance NFRTD, we first focus on the near-field LDOS because a large LDOS usually leads to a large radiative heat flow and vice versa [25]. Our idea is to create a scenario where one side of the diode offers a substantial LDOS contrast as the temperature varies, while the other provides a sharp LDOS peak which serves as a narrow bandpass filter and limits the heat flow to a desired frequency range for ease of control. Such a scenario can be established, for example (Fig. 1), by exploring the phase transitions of MIT materials [26] in conjunction with the SPhPs supported by polar dielectrics [27–29].

The key to maximizing rectification is to maximize the contrast of near-field LDOS across the metal-to-insulator transition. We achieve this by simply using a MIT thin film instead of a bulk, because the low-frequency LDOS of a metal film increases substantially with decreasing film thickness while the opposite often happens for an insulator film, as shown in Fig. 1b-c. Intuitively, as the thickness of a metal film becomes thinner than the skin depth, the surface plasmon polaritons (SPPs) at the two interfaces couple within the film and split into two branches of resonant modes— the low-frequency symmetric and the high-frequency antisymmetric modes [30,31]. Compared to the SPPs on a bulk substrate, the symmetric modes of a thin film shift to higher parallel wavevectors and lead to enhanced LDOS that are thermally accessible [31]. On the contrary, the LDOS for an insulator may decrease over a wide frequency range due to reduced emitting volume as the film thickness decreases [32].

To quantify the effect of film thickness on the LDOS contrast, we derive approximate expressions in the near field of both metals and insulators, assuming nonmagnetic materials so only *p*-polarized surface modes exist. For *p*-polarized evanescent waves, the LDOS at a distance *z* above a suspended film of thickness *t* is given by [25]

$$\rho(\omega,z) = \int_{k_0}^{\infty} \rho_w(\kappa,\omega,z) d\kappa = \int_{k_0}^{\infty} \frac{\kappa^3}{2\pi^2 \omega |\gamma_0|} e^{-2\text{Im}(\gamma_0)z} \text{Im}(R_f) d\kappa, \qquad (1)$$

where $\rho_w$ denotes the wavevector-resolved local density of states (WLDOS) [33,34], $\kappa$ is the parallel wavevector, $k_0 = \frac{\omega}{c}$, $\gamma_0 = \sqrt{k_0^2 - \kappa^2}$, and $R_f$ is the reflection coefficient of the film. Since the near-field heat flow is often dominated by high-$\kappa$ modes with $\kappa \gg k_0$, the WLDOS can be simplified as

$$\rho_w(\kappa,\omega,z) \approx \frac{1}{2\pi^2 \omega} \kappa^2 e^{-2\kappa z} \text{Im}(R_f). \qquad (2)$$

We first consider a film of Drude metal with plasma frequency $\omega_p$ and negligible loss, the permittivity of which can be written as $\varepsilon_m = \varepsilon_m' + i\varepsilon_m''$, where $\varepsilon_m' \approx 1 - \frac{\omega_p^2}{\omega^2}$ and $\varepsilon_m'' \approx 0$. By solving for the poles of $R_f$ at sufficiently low frequencies ($\omega \ll \omega_p$), we find the symmetric resonant modes characterized by $\kappa_r = \frac{2}{|\varepsilon_m'|t}$ (see Supplementary Material for details [35]). The increase of $\kappa_r$ with decreasing film thickness reflects stronger coupling of SPPs within thinner films [30,31]. Upon resonance we have $\text{Im}(R_f) = \frac{|\varepsilon_m'|}{\varepsilon_m''}$, so the WLDOS is written as

$$\rho_{w,m}(\kappa_r,\omega,z) = \frac{2}{\pi^2 \omega \varepsilon''_m |\varepsilon'_m|} e^{\frac{-4z}{|\varepsilon'_m|t}} t^{-2}, \tag{3}$$

with a full width at half maximum of $\Delta\kappa = \frac{4\varepsilon''_m}{(\varepsilon'_m)^2} t^{-1}$. The LDOS of a metal film can then be approximated by

$$\rho_m \approx \rho_{w,m}(\kappa_r,\omega,z)\Delta\kappa = \frac{8}{\pi^2 \omega |\varepsilon'_m|^3} e^{\frac{-4z}{|\varepsilon'_m|t}} t^{-3}, \tag{4}$$

which increases with decreasing film thickness roughly as $t^{-3}$ and peaks at $t = \frac{4z}{3|\varepsilon'_m|}$. Likewise, for an insulator film characterized by $\varepsilon_i = \varepsilon'_i + i\varepsilon''_i$, with low loss and no surface modes, the low-frequency LDOS is derived as [35]

$$\rho_i \approx \frac{27\varepsilon''_i (\varepsilon'^2_i + 1)}{16 e^3 \pi^2 \omega \varepsilon'^2_i} z^{-4} t. \tag{5}$$

Here, the linear scaling with film thickness reflects the linearly decreasing volume. Equations (4) and (5) show that for a hypothetical MIT film, the LDOS contrast between the metal and the insulator phase could potentially scale as $t^{-4}$, in consistency with our intuitive pictures. Further, we plot in Fig. 1b rigorously calculated results assuming desired permittivities, which agree well with the asymptotic analysis.

Guided by the physical insights from ideal materials, we now explore the potential of real MIT materials for large thermal rectification. To this end, we compute the near-field LDOS for vanadium dioxide ($VO_2$)—a prototypical MIT material with a phase transition temperature around 341 K [36-38]. As expected, the LDOS of a 1-nm-thick $VO_2$ film is up to two orders-of-magnitude larger than that of bulk $VO_2$ in the metal phase over a wide spectral range, while a similarly dramatic LDOS decrease is observed in the insulator phase (Fig. 1c). This leads to a substantial increase in the LDOS contrast and forms the basis for large rectification, despite the fact that $VO_2$ has non-negligible loss in both phases and fails to quantitatively follow the asymptotic relation with film thickness (Fig. S1). To form a thermal diode (Fig. 1a), we choose cubic boron nitride (cBN) [39,40] as the polar dielectric that pairs with VO$_2$, because it supports strong SPhPs around $\omega_{cBN} = 2.376 \times 10^{14}$ rad/s where VO$_2$ shows a large metal-to-insulator LDOS ratio (Fig. 1c).

With the materials of our NFRTD selected, we proceed to exact calculations of the forward and reverse heat flow in the parallel-plane configuration. Based on the theoretical framework of fluctuational electrodynamics, the heat flux between two closely spaced bodies across a vacuum gap is given by [41,42]

$$q(T_1,T_2,d) = \int_0^\infty \frac{d\omega}{4\pi^2} \left[\Theta(\omega,T_1) - \Theta(\omega,T_2)\right] \int_0^\infty d\kappa\kappa \left[\tau_s^{12}(\omega,\kappa) + \tau_p^{12}(\omega,\kappa)\right], \tag{6}$$

where $\Theta(\omega,T) = \frac{\hbar\omega}{\exp(\hbar\omega/k_B T)-1}$ is the mean energy of a harmonic oscillator less the zero point contribution at absolute temperature $T$, $d$ is the gap size, and $\tau_s^{12}(\omega,\kappa)$ and $\tau_p^{12}(\omega,\kappa)$ are respectively the transmission probabilities for the $s$- and $p$-polarized modes, which can be expressed in terms of various reflection coefficients ($r$) as follows

$$\tau_{\alpha=s,p}^{12}(\omega,\kappa) = \begin{cases} \dfrac{\left(1-|r_\alpha^1|^2\right)\left(1-|r_\alpha^2|^2\right)}{\left|1-r_\alpha^1 r_\alpha^2 \exp(2i\gamma_0 d)\right|^2}, & \kappa \leq k_0 \\ \dfrac{4\,\text{Im}(r_\alpha^1)\,\text{Im}(r_\alpha^2)\,e^{-2\text{Im}(\gamma_0)d}}{\left|1-r_\alpha^1 r_\alpha^2 \exp(2i\gamma_0 d)\right|^2}, & \kappa > k_0 \end{cases} \tag{7}$$

In Eq. (7), $1 - |r_\alpha|^2$ should be replaced with $1 - |r_\alpha|^2 - |t_\alpha|^2$ for suspended thin films, where $t_\alpha$ is the transmission coefficient.

We show in Fig. 2 the transmission probabilities for three diode configurations including bulk VO$_2$ paired with bulk cBN, VO$_2$ film (10 nm) paired with bulk cBN, and VO$_2$ film (10 nm) paired with cBN film (10 nm), at a gap size of 100 nm. The terminal temperatures of the diodes are fixed at $T_{high}$ = 351 K and $T_{low}$ = 331 K to allow complete phase transition of VO$_2$ [38], which is an isotropic metal at $T_{high}$ in the forward scenario and a uniaxial dielectric when the temperature bias is reversed. Compared to the bulk-bulk case (Fig. 2a-c), the film-bulk case (Fig. 2d-f) shows a larger forward heat flux dominated by the SPhPs within the narrow Reststrahlen band of cBN (Table S1), due to the enhanced low-frequency LDOS of the metallic $VO_2$ film (Fig. 1c). While in the reverse scenario, a much smaller heat flux is observed because the reduced volume of insulating $VO_2$ leads to smaller contributions from the broadband non-surface modes with $\kappa <$ $\text{Re}(\sqrt{\varepsilon_{cBN}})k_0$ (left of the cBN light line in Fig. 2). As a result, a much larger rectification ratio ($R$) of 18.28 is obtained with a $VO_2$ film and bulk cBN than that of two bulks ($R = 0.35$). Moreover, the rectification ratio increases to 24.31 when the bulk cBN is replaced with a thin film (Fig. 2g-i), which leads to smaller LDOS outside the Reststrahlen band and weaker coupling with the insulating $VO_2$ film. This further reduces the reverse heat flux while maintaining a large forward flux. As indicated by the parameter $\eta$ in Table S1 which represents the heat flux mediated by the SPhPs of cBN, creating a scenario characterized by narrowband heat transport is key to achieving large thermal rectification.

By systematically optimizing the film thickness, we predict a maximum rectification of 140.77 by combining a 1-nm-thick $VO_2$ film with a 149-nm-thick cBN film (Fig. 3 and Table S1). The dramatic rectification enhancement primarily originates from the ultrathin $VO_2$ film which features a much larger metal-to-insulator LDOS ratio around $\omega_{cBN}$ (Fig. S2b), consistent with our asymptotic analysis. In comparison, the optimal cBN film is much thicker. This is because the SPhPs will couple within the cBN film if its thickness becomes smaller than the penetration depth ($\approx d = 100$ nm) [29,43], leading to a redshift of the LDOS peak to the frequency range where $VO_2$ shows a smaller LDOS contrast and reduces the rectification (Fig. S2b,d). We also note that when the $VO_2$ film is around 1 nm thick, the rectification ratio remains above 90 regardless of the cBN film thickness, which again reflects the key role of the ultrathin VO₂ film (Fig. 3 and Fig. S2c).

A thermal diode composed of two suspended nanofilms is theoretically appealing but technically challenging. Here, we build upon the optimized suspended-film design and propose a more practical multilayer structure with similarly large rectification ratio (Fig. 4a). First, we add to each film a bulk dielectric substrate with a permittivity $\varepsilon_d$ close to that of vacuum. By setting $\varepsilon_d = 1 + 10^{-4}i$, we obtain a rectification of 4.40 across a vacuum gap of 100 nm, in grim contrast to 140.77 for the suspended-film configuration with the same film thicknesses. This is because of the large reverse heat flux dominated by the broadband propagating modes in the dielectric substrates with $\kappa < \text{Re}(\sqrt{\varepsilon_d})k_0$, which can be effectively suppressed by reducing the thickness of the substrates. By using a 1-μm-thick substrate instead of a bulk, we observe essentially the same spectral characteristics as the suspended-film design and a large rectification ratio of 134.52. We further add a reflective metal layer on the backside of the dielectric substrates to isolate the diode from thermal radiation in the environment. With perfect metal (PM) [44] as the reflector, the spectrum remains almost unaltered and a rectification ratio of 134.81 is achieved. We note that the dielectric substrate should be thicker than the penetration depth ($\approx d$) of the dominating surface modes ($\kappa d \approx 1$) so that any detrimental effect from the metal reflector is minimized [29,43].

Although the above design is promising in principle, we have yet to identify real materials that are appropriate to serve as the dielectric substrates and the metal reflectors. The substrates should have vacuum-like permittivities around $\omega_{cBN}$, and should not induce new surface modes which could increase the reverse heat flow and lower the rectification. We find that potassium chloride (KCl) [45], potassium bromide (KBr) [45], silicon (Si) [46], and SiC [25] represent some of the best candidates. As to the backside reflectors, metals with low loss and high reflectivity such as silver (Ag) and gold (Au) are preferred. In Fig. 4b, we present the rectification ratios as a function of gap size for a few thin-film NFRTDs consisting of different combinations of the candidate materials. All of the diodes feature rectification ratios over 60 at a gap size of 100 nm (Table S2), with a maximum of 78.85 achieved by the pair of $VO_2$/KBr/Ag and cBN/SiC/Ag. The rectification ratio of this particular diode further increases to 105.68 at a 50 nm gap which is readily accessible in the parallel-plane geometry [21,22], and remains greater than 10 even at gap sizes up to ~700 nm. We note that the smaller rectification at larger gaps is due to the rapid decay of the surface modes, while at smaller gaps, the gap-size-dependent penetration depth of the surface modes is responsible. Indeed, thin films can act like bulk materials in the extreme near field and results in rectification ratios approaching the bulk-bulk case [29,43,47]. Last but not least, our thin-film diodes are expected to perform well over a wide temperature range (Fig. S3) and against inevitable film thickness variations in fabrication (Fig. S4).

In summary, we have proposed a mechanism to achieve orders-of-magnitude rectification of radiative heat flow between parallel planes across experimentally attainable vacuum gaps. We combine an ultrathin MIT film with a polar dielectric film, so that the latter enables surface-mode-mediated narrowband heat transport in a spectral range where the former provides a large metal-to-insulator LDOS contrast upon phase transition. Our asymptotic analysis reveals that the LDOS ratio could increase with decreasing film thickness (*t*) as $t^{-4}$, leading to a similarly dramatic rectification enhancement. We demonstrate a rectification ratio of over 140 by pairing thin films of $VO_2$ and cBN at a gap size of 100 nm, which is 400 times better than its bulk counterpart and a 10-fold improvement over state-of-the-art thermal diodes. We have further explored realistic multilayer structures and identified a variety of appropriate materials. Our effort paves the way for the long-awaited use of thermal diodes in solid-state energy and logic devices.

**Acknowledgement**

This work was supported by the Beijing Innovation Center for Engineering Science and Advanced Technology (BIC-ESAT), and the National Natural Science Foundation of China (Grant No. 52076002 and No. 51836004). Q. L. and H. H. contributed equally to this work.

asymptotic and exact LDOS for real materials, the effect of film thickness on rectification, the performance of NFRTDs at different temperatures, Monte Carlo study of diode robustness, and the optimal parameters for different NFRTDs.

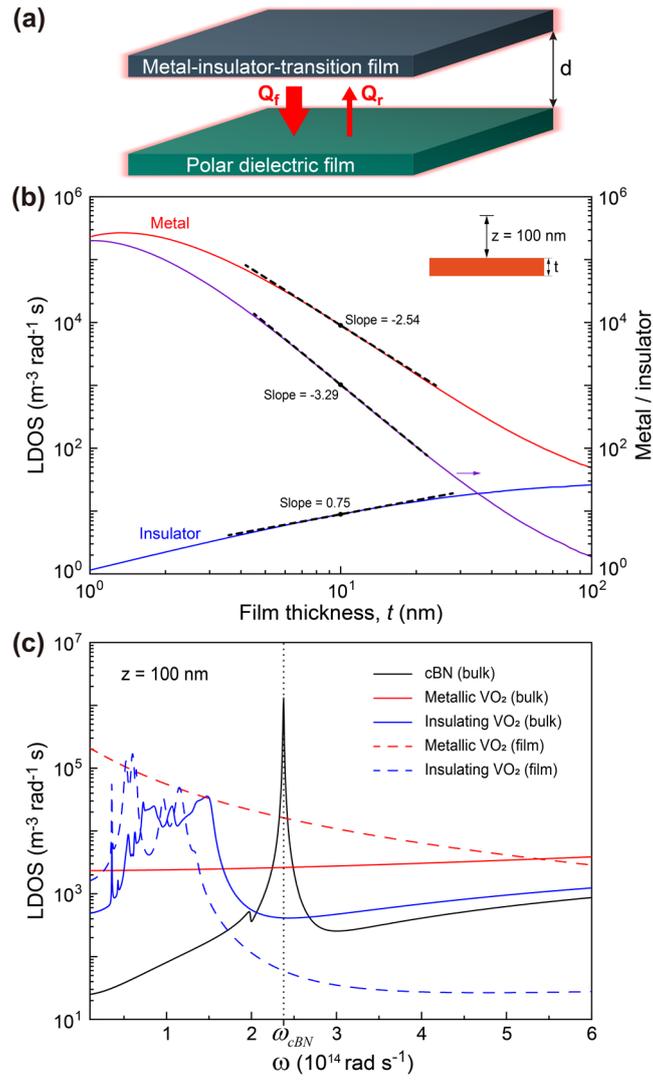

**Fig. 1.** (a) Schematic of the thin-film radiative thermal diode. (b) The LDOS of *p*-polarized evanescent modes 100 nm above a metal and an insulator film and their ratio at $\omega = 10^{14}$ rad/s with varying film thickness. For the metal, we assume $\omega_p = 10\omega$ so $\varepsilon'_m \approx 1 - \frac{\omega_p^2}{\omega^2} = -99$, and set $\varepsilon''_m = 10^{-3}$ to represent a low loss. For the insulator, we set $\varepsilon_i = 2 + 10^{-3}i$. (c) The LDOS above cBN and VO$_2$.

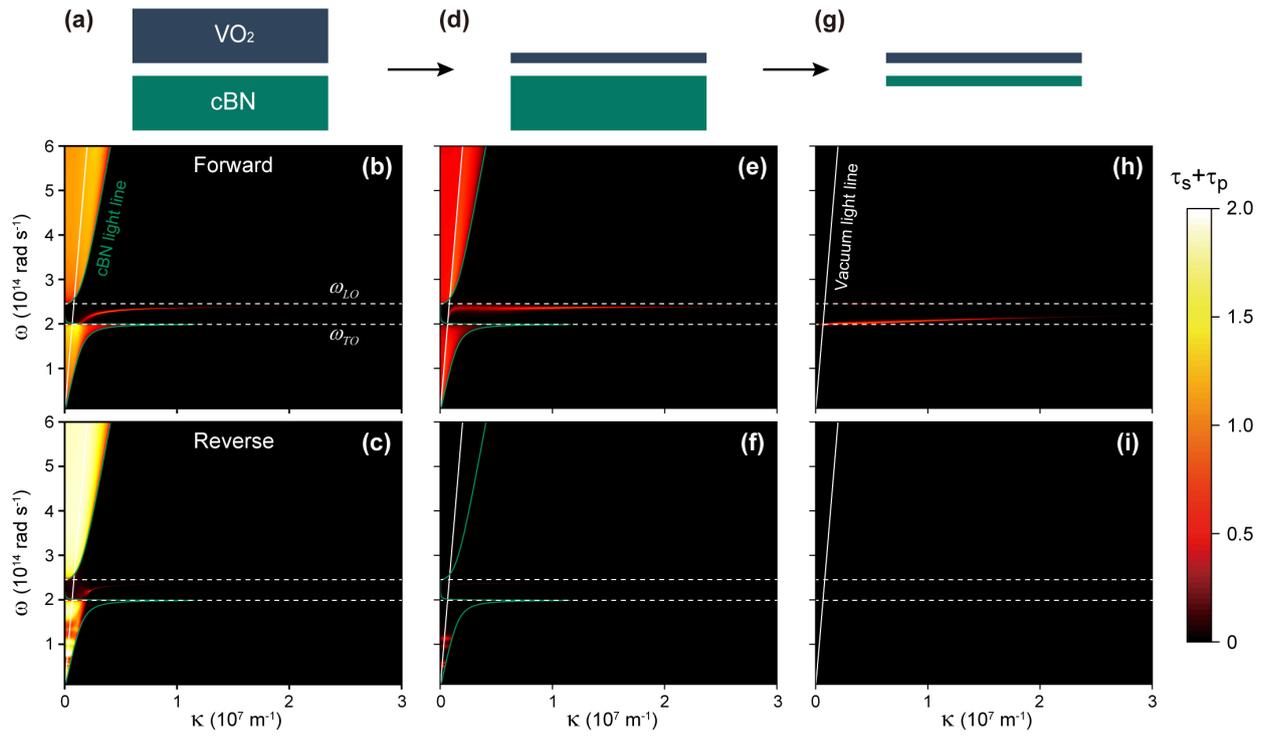

**Fig. 2.** Comparison of the transmission probabilities for bulk and thin-film diodes. (a-c) bulk $VO_2$-bulk cBN, (d-f) $VO_2$ film-bulk cBN, (g-i) $VO_2$ film-cBN film. The vacuum gap is 100 nm wide, and all films are 10 nm thick. The dashed lines mark the zone-center transverse and longitudinal optical phonons ($\omega_{TO}$ and $\omega_{LO}$) in cBN, which enclose the Reststrahlen band.

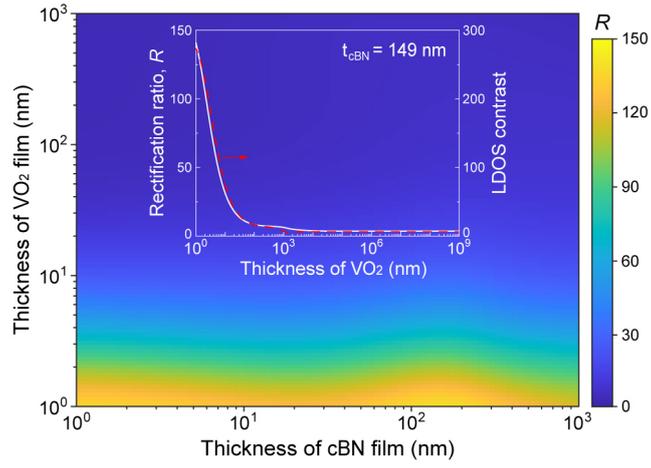

**Fig. 3.** Rectification ratio of a suspended VO$_2$-cBN thin-film diode as a function of film thickness at a 100 nm gap. Inset shows the rectification and the corresponding LDOS contrast at $\omega_{cBN}$ with varying VO$_2$ film thickness, when the cBN film thickness is an optimal value of 149 nm.

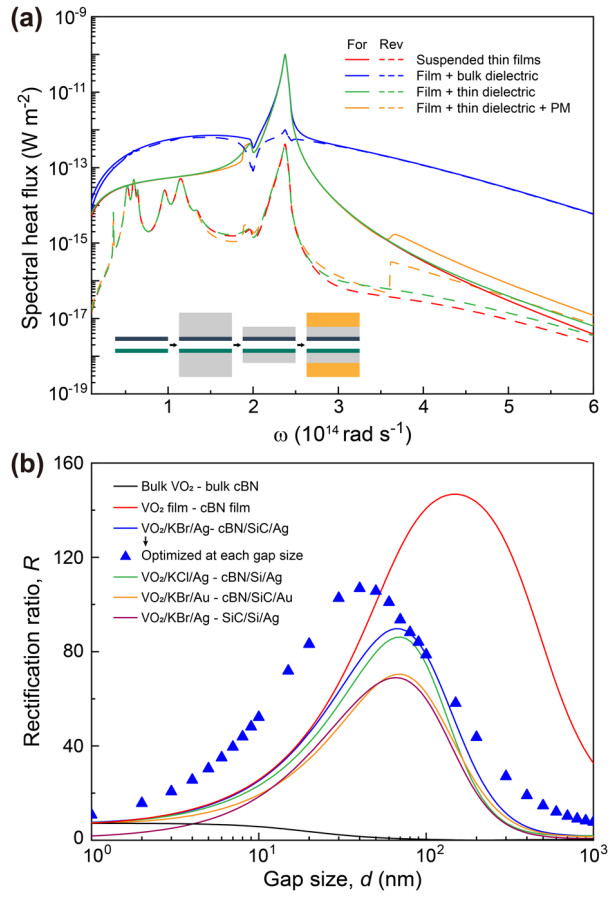

**Fig. 4.** Performance of thin-film diodes with realistic structures and materials. (a) Spectral heat flux in the forward and reverse scenarios across a 100 nm gap. (b) Gap-dependent rectification ratios.